\documentclass{article}
\usepackage{romp33}

\title{Moving frames for cotangent bundles}
\author{Jair Koiller\thanks{\mbox{ } Visiting LNCC, 2000-2001. Permanent position at Funda\c c\~ao Getulio Vargas.
Supported in part by a
PCI-CNPq/Brazil  research fellowship.} \ and \ Pedro de M. Rios\thanks{\mbox{ } Visiting LNCC, 2000-2001, under a PCI/CNPq fellowhip. Present address:  Department of Mathematics, University of California,  Berkeley.}\\
Laborat\'orio Nacional de Computa\c c\~ao
Cient\'{\i}fica \\
Av. Getulio Vargas 333, 25651-070 Petr\'opolis - Brazil\\  jair@lncc.br \ \ prios@math.berkeley.edu \\[2ex]
Kurt M. Ehlers\thanks{\mbox{ } This work was completed on a visit to Rio  supported by FAPERJ/Brazil. }
\\ Department of Mathematics,
Truckee Meadows Community College\\
7000 Dandini Blvd.
Reno NV, 89512-3999 \\ kehlers@scs.unr.edu } 

\begin{document}

\maketitle
\begin{abstract}  Cartan's   moving frames method is a standard tool in  riemannian geometry.   We set up the  machinery for applying   moving frames to cotangent bundles   and  its sub-bundles defined by non-holonomic constraints.  
\end{abstract}


\section{Introduction.}

This paper has a very modest scope: we  present our  ``operational system'' for  Hamiltonian mechanics on cotangent bundles $M = T^*Q$,  based on moving frames. In a related work
\cite{RiosKoiller}, we will present some concrete examples  to convey the algorithmical nature of this formalism.
 
A powerful tool in riemannian
geometry  is the ``method of moving frames'', introduced by \'Elie Cartan.
 Actually, moving frames appeared earlier  in Lagrangian Mechanics:  Poincar\'e
presented in 1901 a moving frame  version of the
Euler-Lagrange equations\footnote{We thank Larry Bates (personal information): moving frames were introduced by Euler \cite{E}.  Certainly moving frames was understood by the caveman who invented the wheel.}, later refered as the ``quasi-coordinates'' method. Cartan himself has advocated applying moving frames in mechanics 
\cite{Cartan}, in particular using his equivalence method. See \cite{JP} for a modern exposition of Cartan's paper. 

When we use a moving frame and its dual coframe, the canonical symplectic
form $\Omega$ in $T^*Q$
deviates from the Darboux format. This is not bad: we use this feature to encode information about the system.

Moving frames are natural  when dealing with Lie groups and with constrained systems, 
either {\it vakonomic} or {\it non-holonomic} (see \cite{Arnold} for background).
Linear  constraints define a distribution $ \mathcal{E}  $ of $s$-dimensional planes   $E_q \subset
T_q Q$, where $Q$ is an $n$-dimensional configuration space, $s < n$.

\section{Basic formalism }

 Let $Q$ be a $n$-dimensional manifold, $TQ$ its tangent  bundle
and $T^*Q$  its cotangent bundle (in both cases we denote by $\pi$ the canonical
projection). Let $  (q_1, ..., q_n)$ be coordinates on
$Q$; the associated  coordinates $p_q = (p,q)$ on
$T^*Q$ are defined by the duality rule $ p_q(\frac{\partial}{\partial q_I}) = 
(p_J dq_J, \frac{\partial}{\partial q_I}) = p_I \,  $
(we adopt the summation convention on repeated indices).
The {\it canonical 1-form} $\omega$ on $T^*Q, \,
\omega
(V_{p_q}) = p_q (\pi_* V_{p_q})
$
writes in the $(p,q)$ coordinates as
$ \omega_{p_q} = p dq$. The {\it canonical symplectic 2-form} 
$\Omega = d \omega$  on $T^*Q$   as 
$\Omega = dp_I \wedge dq_I \,  $.

\subsection{Coframe coordinates for $T^*Q$}

 Let $\{ \epsilon_I = a_{IK} \,dq_K, I=1,...,n\}$   a local coframe in $Q$.
We denote by $\{ e_J  = b_{LJ} \partial/\partial q_L \}$  the dual frame, defined by 
$ \epsilon_I (e_J) = \delta_{IJ}  \,\,.$ 
The matrices $A$ and $B$ are inverses. 
\begin{definition}{Definition} We call quasi-velocities (respectively, quasi momenta) 
the coordinates $(u,q)$ on
$TQ$ (respectively, $(m,q)$ on $T^*Q$) 
defined by 
\begin{equation}  u_q =  u_I \,e_I \ \ , \ \ p_q = m_I \epsilon_I \ \,.
\end{equation}
\end{definition}
The name ``quasi-momenta'' could be replaced without guit my ``momenta''.
For instance, angular momenta  $m_I$  correspond to  $  e_I = {\rm infinitesimal} \,\, {\rm rotations } $ in $\Re^3$. 

Rules of transformation are readily obtained:
\begin{equation}
 p_J = m_I \, \epsilon_I
(\frac{\partial}{\partial q_J}) = m_I a_{IJ} \,\,,\,\,
 m_J = p_I \, dq_I (e_J) = p_I b_{IJ} \,.
\end{equation}

 It is  easy   to write $\omega$ in terms
of the trivialization $ (m,q)$ of $T^*Q$:
\begin{equation} \omega =  p dq = m_I \epsilon_I
\end{equation}
This is the  ``canonical misunderstanding'': the expression $m_I \epsilon_I$   now means a
1-form in $T^*Q$ in coordinates $(m,q)$. The  same
expression (see (\ref{new}))
denotes an element $p_q = m \epsilon(q) = m_I \epsilon_I(q)\, \in T^*_q Q .$
(We use heavier notation when we feel necessary.  We could add  a superscript $\#$
when   thinking of $\epsilon^{\#}_J$    either as a 1-form in Q, or   its
pullback to
$T^*Q$. For the latter a double superscript could be used.   
However, we will try to keep the notation as simple as possible.)

\pagebreak 

The basic idea of this work is to write the canonical 2-form in a non-Darboux format. The following is obvious and will be explored in Theorem \ref{symplecticmoving} further below: 
\begin{theorem}{Theorem} \label{cantwoform}
The canonical 2-form in
$T^*Q$ writes as
\begin{equation} \Omega = d \omega = dp \wedge dq = dm_I \wedge \epsilon_I
+ m_I
d\epsilon_I
\end{equation}
\end{theorem}

\subsection{Earnest  coordinate vectorfields and coframes}

We associate  to the local trivialization
$(m,q)$ such that $ p_q = m_I \epsilon_I$, the lifted coframe for $T^*Q$ given by
\begin{equation}
\{ \epsilon_I \,\,,\,\, dm_I \} \,.
\end{equation}

We will now describe   the  corresponding dual  basis of
 vectorfields in $ T^*Q$.  {\it It turns out that it is not}
$
\{ e_I\, \,,\,\, \partial/\partial m_I \} \,\,.
$
In the correct version, the first set will acquire a fiber component, and will be denoted
$e^*_I$. 

\begin{definition}{Definition} We call   earnest coordinate vectorfields  for $T^*Q$
the coordinate frame associated to  the parametrization $(m,q)$:
\begin{equation}
 X_{q_I}^{\epsilon} ={\frac{\partial}{\partial
q_I}}_{|(m \, {\rm fixed})}\,\,\,\,,\,\,\,\,
\frac{\partial}{\partial m_I} \equiv \epsilon_I \,\,.
\end{equation}
\end{definition}
These vectorfields are dual to the forms $\{ dq_I\,\,,\,\, dm_I \}$, 
differentials of the coordinate functions.
The identification $\partial/\partial m_I \equiv \epsilon_I(q)$,   
  a vertical vectorfield in $T^*Q$, is  the usual
identification of a  vector space with its tangent
space (here, $  T_{p_q} (T^*_qQ) \equiv T^*_qQ$). 

We claim that denoting $  \partial/\partial q_I\,\,\,  $ without subscript,
is misleading. The vectorfields $X^{\epsilon}_{q_I} =  \partial/\partial
q_{I}|_{(m\, {\rm fixed})}$ and
$\partial/\partial q{_I}_{|(p \, {\rm fixed})} $ are different!  
Throughout this work we   reserve  unsubscripted notation
$ \partial/\partial q_I $
for the vectorfield  corresponding to the 
  {\it standard}
coordinates $(p,q)$ for $T^*Q$. Thus if we write $  e_J  = b_{LJ} \partial/\partial q_L  $
thinking as a vectorfield in $T^*Q$, it is assumed  the standard $(p,q)$ parametrization.
 
In fact, we must go back to a standard ``Advanced
Calculus" class.  If  $(q,p)$ and $(q,m)$ are two
sets of coordinates on a fibered manifold, the notation
  $\partial/\partial q_I$ in the two
coordinate systems is  ambiguous:  they differ by a
vertical component\footnote{ Differential forms are  more reliable than vectorfields
in
this regard. Perhaps another
``feminine'' property of forms. Prof. S.S.Chern
insists that forms are of feminine gender,
vectors masculine.}.  
This could be surprising at first sight since the forms
$dq_I$ in the coframes  $\{ dq_I, dp_I \}$ and  $\{ dq_I, dm_I \}$ are the same.  
They are simply
the differentials of the functions
$q_I \circ \pi: T^*Q \rightarrow \Re$
 ($\pi: T^*Q \rightarrow Q$ is the bundle
projection and $q_I: Q \rightarrow \Re$ is the
$I$-th coordinate function).

We introduce matrix notation. We write the (dual) pair frame-coframe  in $Q$  as a line array and column array, respectively:
\begin{equation}  e = (e_1, ... , e_n)
\,\,,\,\,\epsilon =
\left(
\begin{array}{l} \epsilon_1 \\ ...\\ \epsilon_n
\end{array}
\right) \,\,\,\,\,\,\,(  \epsilon \cdot e  =  I_n .) 
\end{equation}
 Write
\begin{equation} \label{A} \epsilon_I = a_{IJ} dq_J\,,
{\rm that}\,{\rm is},\,\,
 \epsilon = A dq\,\,,\,\, 
A = (a_{IJ})\,\,,
\end{equation}
and we recall
\begin{equation}
(e_1, ..., e_n) =
(\partial/\partial q_1, ... ,
\partial/\partial q_n) B\,\,,\,\,  B = A^{-1} \,\,.
\end{equation} 
Then
$m_I
\epsilon_I = p_J dq_J 
$  implies (as we already saw)
$  p_J = m_I a_{IJ} \,\,.$
\begin{lemma}{Lemma}
 (The importance of being earnest)
Assume $\epsilon$ and $dq$ related by (\ref{A}). The corresponding
coframes in $T^*Q$ are related by
\begin{equation} \label{basic}
\left( \begin{array}{l} dq_I \\ dp_I
\end{array}
\right) = \left( \begin{array}{ll}  I & 0 \\
                              \Lambda      &
A^{\dagger}  
\end{array}
\right)
\left(
\begin{array}{l} dq_J\\ dm_J
\end{array}
\right) \,\,,
\end{equation}
where
\begin{equation}
\Lambda_{IJ} =  m_K
\partial a_{KI}/\partial q_J  \,\,.
\end{equation}
The corresponding dual frames in $T^*Q$ are  related by
\begin{equation}
\left(X^{\epsilon}_{q_J}  \,\,\, \partial/\partial
m_J
\right) = \left( \partial/\partial q_I \,\,\,
\partial/\partial p_I      \right) \,\,.\left(
\begin{array}{ll}  I & 0 \\
                                     \Lambda &
A^{\dagger}  
\end{array}
\right)
\end{equation}
\end{lemma}
Explicitly,
\begin{equation}
X^{\epsilon}_{q_J} = \partial/\partial q_J + 
m_K
(\partial a_{KI}/\partial q_J)
\,\,\frac{\partial}{\partial p_J}
\,\,\,,\,\,\,
\partial/\partial m_J =  a_{JI} \partial/\partial
p_I  \,\,.
\end{equation}
Summarizing: the vectorfields $X^{\epsilon}_{q_I}$ and
$\partial/\partial q_I $ are different. 
However, their difference is a vertical vectorfield,  their projections over $TQ$ by   $\pi_* : T(T^*Q)
\rightarrow TQ$  coincide. We say that $X^{\epsilon}_{q_I}$ acquires a {\it spiritual
component} relative to the standard coordinates $(p,q)$.

\subsection{Extended frame $\{ e^*_I, \frac{\partial}{\partial m_I} \}$
  for $ T(T^*Q)$ and coframe $\{ \epsilon_I \,,\, dm_I \} $ for $ T^*(T^*Q)$.}

 We now change the first part of the
 the coordinate basis, $X^{\epsilon}_{q_I}$,  to vectors $e_I^*$. The superscript $^*$   is a reminder
that   $ e_I^* \in T(T^*Q)$, not to $TQ$ and also a reminder that
it has a spiritual component.
 We get after a simple computation
\begin{lemma}{Lemma}
\begin{equation}
\left( \begin{array}{l} dq_I \\ dp_I
\end{array}
\right) = \left( \begin{array}{ll}   B &
0
\\
 \Lambda B & A^{\dagger}  
\end{array}
\right)
\left(
\begin{array}{l} \epsilon_L\\ dm_L
\end{array}
\right) \,\,,
\end{equation}
Dualizing, we get
\begin{equation}
\left( e_J^*  \,\,\, \partial/\partial m_J
\right) = \left( \partial/\partial q_I \,\,\,
\partial/\partial p_I\right) \,\,.\left(
\begin{array}{ll}   B & 0
\\
 \Lambda B & A^{\dagger}  
\end{array}
\right)
\end{equation}
\end{lemma}

In short, the transformation rules to the moving
frame in $T^*Q$ are given by {\rm (in shorthand notation)}, in terms of  the
standard coordinates $(p,q)$:

\begin{equation} \label{e*} e^* = \partial/\partial q B  +
\partial/\partial p \Lambda B =  e  + 
\partial/\partial p \Lambda B\,,\,\,
\end{equation}
\begin{equation}  \label{u}
\partial/\partial m =
\partial/\partial p A^{\dagger} \,\,\, \,\,\,( \,\,
\,\,\, {\rm equivalently}\,\,\,  
\epsilon = A   dq \,\, )
\end{equation}
The last equality is due to the identification 
$  \,\, \partial/\partial m_I = \epsilon_I \,\,,\,\, \partial/\partial p_I = dq_I \,\,. $  
The extended moving coframe in $T_{p_q}^*(T^*Q)$ is
$\epsilon_I \,, \,\,  dm_I \,\,$,  dual to 
  $e^*_I, \partial/\partial m_I   \in
T_{p_q}(T^*Q)$.

Notice  the importance of being earnest: the frames $ \{ e_I \}$ and $\{ \epsilon_J \}$
are dual in  $V = T_qQ, V^* = T_q^*Q$. The frames $ \{ \frac{\partial}{\partial m_I} \}$ 
and $\{ dm_J \}$
are dual in $W = T^*_qQ, W^* = (T_q^*Q)^*$,   but $ \{ e_I, \frac{\partial}{\partial m_I} \}$ and $\{ \epsilon_J, dm_J \} $ are {\it NOT} dual in  $T_{(p,q)} T^*Q, T^*_{(p,q)} T^*Q$.
The basic reason is that $T_{(p,q)} T^*Q \neq  T_qQ \times T_q^*Q $. 
Shortly we will give the precise formula for $e^*_I$. 

\section{Symplectic form in    
$ \{e^*, \partial/\partial m \} $  and Poisson brackets in   $\{\epsilon, dm \}$}

After this quite dull preparation, we are finally  able
to write down  a more interesting formula: 

\begin{theorem}{Theorem} \label{symplecticmoving}
In the basis  $ \{e^*, \partial/\partial m \} $, the
canonical symplectic form
$ \Omega = dp \wedge dq = dm_I \wedge \epsilon_I^{\#} +  m_I d\epsilon^{\#}_I $
writes  as
\begin{equation} [\Omega]_{\{e^*,\partial/\partial m \}} = \left(
\begin{array}{ll}  E & -I\\   I & \,\,\,\,0
\end{array} \right)
\end{equation}
with
\begin{equation} \label{R}
 E_{JK} = m_I d\epsilon_I
(e_J,e_K) = - m_I
\epsilon_I[e_J,e_K]
\end{equation}
\end{theorem}
\noindent{\bf Proof}.  We use  Theorem \ref{cantwoform} and Cartan's magic formula for differentiating 1-forms\footnote{Cartan's formula is the deepest fact used in this paper.}. By duality, the first term $dm_I
\wedge
\epsilon_I^{\#}$ yields a familiar matrix:
$$ \left( \begin{array}{ll}   0 & -I\\   I &
\,\,\,\,0
\end{array} \right) \,\,\,.
$$
The ``magnetic block'' $E$ ($E$ for Euler) results from employing
Cartan's formula:
\begin{eqnarray}  d\epsilon^{\#}_I (e^*_J,e^*_K)  
  \,\,({\rm in}\,\, T^*Q) & = & d\epsilon_I (e
_J,e_K)  \,\,({\rm in}\,\,Q) =  \nonumber \\
 & = & e_J
\epsilon_I(e_K) - e_K \epsilon_I(e_J) - \epsilon_I
[e_J,e_K]
\end{eqnarray}
and we observe that the first two terms vanish. \rule{5pt}{5pt}

As the Poisson structure is a skew-symmetric  tensor
of type $(0,2)$, it operates on two
elements of $ T^*_{p_q}(T^*Q)$. It is natural
to use the basis  $ \{
\epsilon_I \,, \,\,  dm_I \,\, \}$.

\begin{theorem}{Theorem} The Poisson
bracket matrix relative to 
$\, \epsilon_I \,, \,\,  dm_I \,\,$ is
\begin{equation} \label{poisson} [\Omega]^{-1} = [\Lambda]
=
\left(
\begin{array}{ll}   0_n & I_n\\   -I_n & \,\,\,\,E
\end{array} \right)
\end{equation}
Equivalently,
\begin{equation}
\Lambda = \sum_I \,\, e^*_I \wedge \frac{\partial}{\partial{m_I}} + \frac{1}{2}\,\,\sum_{1 \leq J,K \leq n} \, E_{JK} \,
\frac{\partial}{\partial{m_J}} \wedge \frac{\partial}{\partial{m_K}} \ . 
\end{equation}
\end{theorem}

 We now observe that
$$
\Lambda   =   \sum_I \,\, e^*_I \wedge \frac{\partial}{\partial{m_I}} - \frac{1}{2}\,\,  E_{IJ} \,
\frac{\partial}{\partial{m_J}} \wedge \frac{\partial}{\partial{m_I}} = 
  \tilde{e}_I \wedge \frac{\partial}{\partial{m_I}}
$$
where
$$ \tilde{e}_I = e^*_I(q) -
\frac{1}{2} E_{IJ}
\frac{\partial}{\partial m_J} \,\,\,.
$$

The $(0,2)$ Poisson tensor can also be written as
\begin{equation}
\Lambda = \sum_I \, \frac{\partial}{\partial q_I} \wedge \frac{\partial}{\partial p_I}
= \sum_I \, e_I \wedge \frac{\partial}{\partial m_I}
\end{equation} 
 The last equality is   ridiculous. As 
$$  e_I = b_{JI}  \partial/\partial q_J \,\,,\,\,  \frac{\partial}{\partial m_I} = \frac{\partial}{\partial p_J} \, a_{IJ} \,\, \,\,\,({\rm see}\,\, (\ref{u}))
$$
and since $A = B^{-1}$ we have
$$ \sum_I \, e_I \wedge \frac{\partial}{\partial m_I} =  
b_{JI} \frac{ \partial}{\partial q_J } \, \wedge \frac{\partial}{\partial p_K} \, a_{KI} = a_{IK}\,b_{JI}\,\frac{ \partial}{\partial q_J }  \, \wedge \frac{\partial}{\partial p_K} = \delta_{KJ} \, \frac{ \partial}{\partial q_J }  \, \wedge \frac{\partial}{\partial p_K} = \frac{ \partial}{\partial q_J }  \, \wedge \frac{\partial}{\partial p_J}.
$$

Thus one could guess that $\tilde{e}_I = e_I$, but in fact brute force computation gives:
\begin{equation}
  e^*_I(q) = e_I + m_K\frac{\partial a_{KL}}{\partial q_R}(b_{RI}b_{LJ} + b_{RJ}b_{LI})\frac{\partial}{\partial m_J} \,\,\,.
\end{equation}

This gives the expression for the spiritual component, as promised before. Note that the second term does not contribute when wedging with $\frac{\partial}{\partial m_I}$ and performing the summation. 

\section{Examples}

\subsection{Lie groups and KAKS bracket. }  \label{KAKSsection}
  
Let the configuration space be a Lie group $Q = G$ , 
$ e_I$ and $\epsilon_I$ dual left-invariant vectorfields
and forms, with structure constants defined by
$[e_J,e_K] = c^I_{JK}\,e_I \,.$  Then
\begin{equation}  E_{JK} = m_I d\epsilon_I
(e_J,e_K) = - m_I
\epsilon_I[e_J,e_K] = - m_I c^I_{JK}
\end{equation}
does not depend on $g \in G$.
Write
$$   V_{p_g}^a = X_J^a e_J^* + z_J^a \epsilon_J\,\in T_{p_g}(T^*G),\,\,\, a=1,2 $$
so  
\begin{equation} \label{KAKS}
\Omega(V^1,V^2) = (X^1, z^1)  \left(
\begin{array}{ll}   E & -I\\   I & \,\,\,\,0
\end{array} \right) \, \left(
\begin{array}{l}   X^2  \\ z^2 \end{array} \right) = X^2 z^1 - X^1 z^2 + X^1 E X^2\,\,.
\end{equation}
We denote  $X_J^a \, e_J(\rm{id}) = L_{g^{-1}}(\pi_* \, V_{p_g}^a) $ simply as $X^a \in \mathcal{G}$
and therefore
\begin{equation} \label{KAKS1} X^1 E X^2 = - m_I \epsilon_I(g) [X_1,X_2]_g = - (L_g)^* (p_g)[X^{left}_1,X^{left}_2]_{\rm{id}} \,\,.
\end{equation}

What if we replace left by right-invariant vectorfields $f_I$ and forms $\theta_I$?
The basic formula stays the same:
$$ \Omega(V_1,V_2)  = X^2 z^1 - X^1 z^2 + X^1 E X^2\,,
$$
but now $X_J^a \, f_J(\rm{id}) = R_{g^{-1}}(\pi_* \, V_{p_g}^a) $
and
$$  X^1 R X^2 = - m_I \theta_I(g) [X_1,X_2]_g = - (R_g)_* (p_g)[X^{right}_1,X^{right}_2]_e \,\,.
$$
where
\begin{equation}  \label{right}
[X^{right}_1,X^{right}_2]_e = - c^I_{JK}\, X^1_J X^2_K \,f_I  \,.
\end{equation}
Notice the extra minus sign arising from the Lie bracket structure.   Here we used
the well known Lie-group fact:
 {\it if one extends vectors in} $\mathcal{G}$ {\it right invariantly the
structure coefficients in the Lie bracket appear with opposite sign.}

Equations (\ref{KAKS}) and (\ref{KAKS1}) lead to the   KAKS   (Kirillov-Arnold-Kostant-Souriau)  bracket in the dual  
$\mathcal{G}^*$ of the Lie algebra \footnote{This bracket was found independently by S. Lie \cite{Weinstein}.} . 

The commutation relations for the forms $\epsilon_I \,,\,  dm_I \,\,$ in $T_{u \cdot \epsilon}^*(T^*G)$ are given by
\begin{equation}
\left\{\epsilon_I, \epsilon_J \right\} = 0\,,\, \left\{dm_I, \epsilon_J \right\} = \delta_{IJ}\,,\,
\left\{dm_I, dm_J \right\} = E_{IJ} = - m_K c^K_{IJ}
\end{equation}
The last commutation  formula implies  for  $f,g: \mathcal{G}^* \rightarrow \Re$, that
at $\mu \in \mathcal{G}^*$,
\begin{equation}
\left\{ f\,,\, g \right\}(\mu) = - \mu [\frac{\delta}{\delta f},\frac{\delta}{\delta g}] 
\end{equation}
where  $df(\mu) \in 
T_{\mu}^* \mathcal{G}^*\,$
is identified with $ \frac{\delta}{\delta f}(\mu) \in \mathcal{G}$.

\subsection{Principal bundles with connection} \label{principalbundles}

We use heretofore  the
following convention: capital roman letters
$I,J,K,$ etc., run from
$1$ to
$n$.
 Lower case roman characters $i,j,k $ run from $1$ to $s$.  Greek
characters $\alpha, \beta, \gamma$, etc., run from  $s+1$ to $n$.

Let $\pi: Q^n \rightarrow S^s$ a principal bundle
with Lie group $G^r$, where  $ r = n-s$. For definiteness, we take $G$ acting on the left.
Fix a connection $\lambda = \lambda(q): T_q Q \rightarrow \mathcal{G}$ defining a
$G$-invariant distribution $\mathcal{E}$ of horizontal subspaces.
Denote by  $K(q) = d\lambda \circ {\rm Hor}: T_q Q \times T_q Q \rightarrow
\mathcal{G}$ the curvature 2-form (which is, as
well known,
$Ad$-equivariant).

Choose a local frame $\overline{e}_i$ on $S$.  
For simplicity, we may assume that
\begin{equation} \label{easychoice}
\overline{e}_i = \partial/\partial
s_i
\end{equation} are the coordinate vectorfields of a
chart
$s: S
\rightarrow \Re^s$.

Let  $e_i = h(\overline{e}_i)$ their horizontal
lift to $Q$.  We complete to a moving frame
of $Q$ with vertical vectors $e_{\alpha}$
which we will specify in a moment.
The dual basis will be denoted $\epsilon_i, 
\epsilon_{\alpha}$ and we write $ p_q = m_i \epsilon_i + m_{\alpha}
\epsilon_{\alpha}$. These are in a sense the ``lesser moving'' among all the moving frames
adapted to this structure.
We now describe how the $n \times n$ matrix $ E =
(E_{IJ})$ looks like in this setting.\\

i) The  $s \times s$ block  $(E_{ij})$.\\

Decompose 
$[e_i,e_j] =
h[\overline{e}_i,\overline{e}_j] + V[e_i,e_j] =
V[e_i,e_j]
$ 
into
vertical and horizontal parts.  The choice
(\ref{easychoice}) is convenient, since $ \overline{e}_i$ and $\overline{e}_j$  commute:
$[e_i,e_j]$ is vertical. Hence
 \begin{equation} E_{ij} =  - p_q[e_i,e_j] = - m_{\alpha}
\epsilon_{\alpha} [e_i,e_j] \,\,.
\end{equation}

Now by Cartan's rule,  
$$ K(e_i,e_j) = e_i \lambda(e_j) - e_j
\lambda(e_i) -
\lambda [e_i,e_j] =  -
\lambda [e_i,e_j]
\in
\mathcal{G}
$$
Thus we have shown that  
\begin{equation} [e_i,e_j]_q = - K(e_i,e_j) \cdot q
\end{equation}
Moreover, let
$ J: T^*Q \rightarrow \mathcal{G}^* $ be the momentum mapping. We have
$$ (J(p_q), K_q(e_i,e_j) ) = p_q \,(
K(e_i,e_j).q\,) = - p_q [e_i,e_j] \,\,\,\,\,\,(= E_{ij})
$$
\begin{theorem}{Theorem} (The  J.K formula)
\begin{equation}  E_{ij} = (J(p_q), K_q(e_i,e_j) )  
\end{equation}
\end{theorem}
This gives a nice   description for   this block, under the choice
$[\overline{e}_i,\overline{e}_j] = 0.$
Notice that the functions
$E_{ij}$ depend on $s$ and the components
$m_{\alpha}$, but do not depend on $g$. This is
because the $Ad^*$-ambiguity of the momentum
mapping $J$ is cancelled by the $Ad$-ambiguity
of the curvature $K$.\\

ii) The  $ r \times r$ block  $(E_{\alpha \beta})$. \\

Choose a basis $X_{\alpha}$ for $ \mathcal{G}$. We take   $e_{\alpha}(q) = X_{\alpha} \cdot q$  as the  vertical distribution.  Choosing a point $q_o$ allows
identifying  the Lie group $G$ with the fiber containing $G q_o$, so that ${\rm id} \mapsto q_o$.
Through the mapping $g \in G \mapsto gq_o \in Gq_o$ the vectorfied $e_{\alpha}$ is identified
to a {\it right} (not left!) invariant vectorfield in $G$.  Thus the commutation relations
for the $e_{\alpha}$ are as in (\ref{right}) so that
$  [e_{\alpha},e_{\beta}] = - c_{\alpha \beta}^{\gamma} \, e_{\gamma}
$ 
appears with a minus sign. Therefore
\begin{equation}
E_{\alpha \beta} = m_{\gamma} c_{\alpha \beta}^{\gamma} \,\,. 
\end{equation}

iii) The  $s \times n$ block $(E_{i \alpha})$.  \\

The vectors  $[e_i, e_{\alpha}]$ are vertical, but their values depend on the
 specific  principal bundle one is working with.  Given a section $\sigma: U_S \rightarrow Q$ over the
coordinate chart  $s: U_S \rightarrow \Re^m$ on $S$, we need to know the coefficients
$b^{\gamma}_{i \alpha}$ in the expansion
$$ [e_i, e_{\alpha}](\sigma(s)) = b^{\gamma}_{i \alpha} (s) \, e_{\gamma} \,\,.$$
Then
\begin{equation}
 E_{i \alpha}(\sigma(s))  = - m_{\gamma}\, b^{\gamma}_{i \alpha} (s) \,\,.
\end{equation}
At another point on the fiber, we need the adjoint representation  $Ad_g: \mathcal{G}
\rightarrow \mathcal{G}, \, X \mapsto g_*^{-1} X g, \,$  described by a matrix
$(A_{\mu \alpha}(g))$ such that
\begin{equation} Ad_g (X_{\alpha}) =  A_{\mu \alpha}(g) X_{\mu} \,\,\,.
\end{equation}
Then
\begin{equation}
[e_i, e_{\alpha}](g \cdot \sigma(s)) = - m_{\gamma} b^{\gamma}_{i \mu} (s) A_{\mu \alpha}(g) \,\,.
\end{equation}

\section{Nonholonomic mechanics}

Consider the Lagrange-D'Alembert
equations    
\begin{equation} \label{DAlembert}
X_L^{D'A}\,:\,\,\, \frac{d}{dt}\, \partial L/\partial \dot{q} - \partial L/\partial q = \lambda A\,\,,\,\, A \dot{q} = 0,
\end{equation}
with $q \in \Re^n, \lambda \in \Re^r, \,  A(q)$ a $r \times n$  matrix. For the regularity assumptions, see \cite{SCS}. 
 
 More intrisically, the constraint equations define
a $ s = n - r$ dimensional distribution $\mathcal{E} $ of subspaces $E_q \subset T_qQ$.
The constraint forces $\lambda A \in T_q^* Q$ belong to the annihilator $\mathcal{E}^{\perp}$,
of $\mathcal{E}$. This is a distribution of $r$-dimensional subspaces $E_q^{\perp} \subset T^*_q Q$.
Under the   Legendre transformation ${\rm Leg}: TQ \rightarrow T^*Q ,\,\,  p = \frac{\partial L}{\partial \dot{q}}\,\,,\,\,\, L + H = p \cdot \dot{q} $ the Lagrange-D'Alembert system (\ref{DAlembert}) of equations
 $  (q,\dot{q}) \in \mathcal{E} \mapsto X_L^{D'A}(q,\dot{q}) \in T \mathcal{E}$ 
 transforms into
the vectorfield  
$ (q,p) \in {\rm Leg}(\mathcal{E}) \mapsto X_H^{D'A}(q,\dot{q}) \in T {\rm Leg}(\mathcal{E})$ 
given by the differential-algebraic system
\begin{equation} \label{constrsymp}
\Omega(X^{D'A}_H + \lambda, \bullet) = - dH(\bullet),
\,\,\, \bullet \in T(T^*Q) \,\,\,,\,\,\,
 \lambda \in E^{\perp} \,\, ,\, \,
 \pi_* X^{D'A}_H (q,p) \,\,
\in E_q
\end{equation}
 where   $\pi: T^*Q
\rightarrow Q$ is the bundle projection.
Here  we identify the the constraint forces  (semibasic vectors)
$\lambda(p,q) \in T_{p_q}(T^*_q Q) \subset T(T^*Q)$ as   elements of
$E_q^{\perp} \subset T^*Q $. 

 The ODEs  (\ref{DAlembert}) restricted to $(q,\dot{q}) \in \mathcal{E}$  must satisfy  $X_L^{D'A}(q,\dot{q}) \in T \mathcal{E} $ (self-consistency requirement).
The process of ``eliminating the multiplier'' $\lambda$
involves differentiating the condition $A(q) \dot{q} = 0$. In other words, self-consistency is precisely what is used to {\it construct} the system of ODEs, ``eliminating the multipliers'' $\lambda$. This step involves differentiating the condition $A(q)\dot{q} = 0$.

\subsection{Equations of motion.}

Consider an {\it  adapted frame}  $e_i,
e_{\alpha}$  to
$\mathcal{E}$ (this  means
that  $e_i(q) \in E_q$) and its dual coframe
$\epsilon_i, \epsilon_{\alpha}$.
 Notice that we are not assuming  that
$e_{\alpha}$ are orthogonal to $E_q$ with respect to a given metric. In fact, when $H$ comes from
a natural Lagrangian $L = T - V$, it seems   natural to
choose  $e_I$ orthonormal with respect to $T$, as proposed by Cartan \cite{Cartan}.
However 
in the presence of symmetries transversal to the constraints, it may  be more interesting to choose  the
$e_{\alpha}$ as vectorfields generated by the symmetries \cite{Koiller}. See section \ref{reduction} below.

Our approach emphasizes the Lie brackets of the frame vectorfields, but some authors prefer to compute the almost Poisson bracket entirely within the bracket formalism using suitable projections, see \cite{B}.  
We write in full the
defining equation (\ref{constrsymp}):
\begin{equation}
\Omega (v_j e^*_j + \dot{m}_J \partial/\partial m_J +
\lambda_{\alpha}
\partial/\partial m_{\alpha}\,,\,  A_I e^*_I + B_I \partial/\partial m_I) = -
dH (A_I e^*_I + B_I
\, \partial/\partial m_I)\,\,\,,
$$
$$  X_H^{D'A}  =  v_j e^*_j + \dot{m}_J \partial/\partial m_J \,,
\,  \lambda = \lambda_{\alpha} \partial/\partial m_{\alpha}
\,,\,\,
\bullet = A_I e^*_I + B_I \partial/\partial m_I \,\,.
\end{equation}
Here the superscript ``D'A''  stands for constrained Lagrange-D'Alembert, not to be
confused with constrained variational type \cite{Arnold}. Using Theorem \ref{symplecticmoving} we get
\begin{equation}
 - v_k B_k + \lambda_{\alpha} A_{\alpha} + \dot{m}_J \, A_J + v_j E_{jI} A_I =
- A_R dH(e^*_R) - B_S \frac{\partial H}{\partial m_S} 
\end{equation}

Equating the coefficients of $A_R$ and $B_S$ we obtain
the    equations for nonholonomic systems.
 First  notice that   in the left hand side there are no terms with $B_{\alpha}$, hence
we are {\it forced} to work in the subset $P$  of
$T^*Q$ given by
$ \frac{\partial H}{\partial m_{\alpha}} =
0\,\,,\,\,\,
\alpha = s+1, ..., n
\,\,.
$
\begin{theorem}{Theorem}  \label{nonhrules}  An ``Operational System'' for nonholonomic systems:\\
\\
(i) The condition 
\begin{equation} \frac{\partial H}{\partial m_{\alpha}} =
0\,\,,\, \alpha = s+1, ..., n 
\end{equation} 
is   equivalent to  $ P = {\rm Leg } (\mathcal{E})$, where
$ {\rm Leg}: TQ \rightarrow T^*Q$ is the Legendre transformation. 
Assume the hypotesis for the implicit
function theorem  ($P$ intersects $\mathcal{E}^{\perp}$ transversaly)  so we can
solve for the
$m_{\alpha} = m_{\alpha}(q, m_k)$ in terms of the $n + s$ variables $q,
m_k$.\\
\\
(ii) The dynamic equations are given
by:
\begin{equation} v_i =
\frac{\partial H}{\partial m_{i}}\,\,,\,\, \dot{m}_i +  v_k \, E_{ki}
 = - dH_{|(q,m)}(e^*_i)
\end{equation}
where for $m = (m_i, m_{\alpha})$ the   $m_{\alpha}$ are as in (i).\\
\\
(iii) The multipliers are
explicitly given by
\begin{equation}
\lambda_{\alpha} = -
\dot{u}_{\alpha} 
- v_j E_{j \alpha} -
dH_{|(q,m)}(e^*_{\alpha})
\end{equation}
\end{theorem}

 In practice, the reader should not fear having difficulties in computing
$ dH_{|(q,m)}(e^*_i)$.  Recall the earnest duality $\{e^*_I, \partial/\partial m_J \}$
to $\{ \epsilon_K, dm_L \}$, so it sufficies to write
\begin{equation}
 dH = \alpha_I \epsilon_I + \beta_J dm_J,
\end{equation}  so 
$  dH(e^*_I) = \alpha_I \,\,,\,\,
dH(\partial/\partial m_J) = \beta_J .$  

 The standard approach to eliminate the
 constraints $\lambda $ in (\ref{DAlembert})
requires {\it differentiating} the constraint equations $A
\dot{q} = 0\,\,$.  The symplectic approach
  seems to be merely an algebraic calculation, but this is not the case. Differentiation is automatically built 
in the algebra since we differentiate the $\epsilon_I$. Equivalently, the almost Poisson bracket approach, first introduced by  van der Schaft
$\&$ Maschke \cite{SM},   also requires a differentiation, namely
taking the Lie bracket of vectorfields satisfying  the constraint
equations. 

\subsection{Reduction} \label{reduction}
Identify a point  of
$P$ with its coordinates $(q,m_k)$.
Therefore, in order to compute the
$ (n+s) \times (n+s)$ (almost)-Poisson matrix,
with respect to the basis $\epsilon_I,dm_k$ it
sufficies to cut the last $ r = n-s $ rows and columns
of $[\Lambda]$ in (\ref{poisson}).
This gives
\begin{equation}  \label{prepoisson}   [\Lambda]_{constrained} = \left(
\begin{array}{lll}   0_{s \times s} & 0_{s \times r}
& I_{s \times s}  \\ 0_{ r
\times s} & 0_{r \times r} & 0_{r
\times s} \\   -I_{s \times s} &  0_{s \times r}  &
 E^c
\end{array} \right)
\end{equation}
where
\begin{equation} \label{Rc} E^c_{jk} =   -
 p_q \cdot [e_j,e_k]  \,\,,\,\, j,k = 1, ..., s
\end{equation}
and
$ p_q \in P \subset T^*Q$ is the point
with coordinates $q, m_k, m_{\alpha}$ satisfying 

$$ m_{\alpha} =
m_{\alpha}(q,m_k)\,\,.$$ 

 Notice  that the middle rows and columns vanish. 
In the presence of transversal
symmetries yielding a principal bundle $G^r \hookrightarrow Q^n \rightarrow S^s$, 
we can ``zip'' (compress) the system down to an almost Poisson structure in $T^*S$. Let
$ H^*(q,m_i) = H(q,m_i, m_{\alpha}(q,m_i))$.  Since $\partial H/\partial m_{\alpha} =0$,
we have $\partial H^*/\partial q =  \partial H/\partial q\,\, , \,\,
\partial H^*/\partial m_i =  \partial H/\partial m_i$ so the right hand side in
Theorem \ref{nonhrules} is preserved under reduction. 

In many nonholonomic systems such as a rigid convex body rolling on a flat plane, the symmetry group does indeed 
intersect the constraints transversally.
Internal symmetries (that is,
satisfying the constraints) will produce
conserved quantities \cite{Arnold} and the quest for integrability of the reduced system. 

We have observed in {\it very simple} examples
\cite{RiosKoiller} that the compressed system is sometimes {\it conformally} symplectic. 
However, further work on these issues indicate that such a property is far from being the rule. 
We will report on this work elsewhere.

\subsection{Final remarks}

Local symplectic geometry is considered to be trivial due to  Darboux theorem. Global symplectic geometry is reputed to be difficult\footnote{Recently however, H. Hofer proposed
introducing piecewise linear
symplectic structures as a way to pass from local to global.}. We believe that moving
frames can  be useful for studying  manifolds  endowed with a distinguished skew-symmetric structure   
(symplectic, Poisson, Dirac, Jacobi, quasi-Poisson, almost-Poisson...) together with some completing structure 
(homogeneous, riemannian, Kahler...), for which the Darboux charts could become cumbersome.  

Also, ODEs for nonholonomic  systems have been  derived  again and again, but the main question remains open:
to  construct a   theory for nonholonomic systems,  similar to that Hamilton  and Jacobi created for holonomic systems. 
In future work we will present some ideas on the  issues of 
symmetry, reduction and integrability.  Here we just present two simple observations to conclude this paper \footnote{These remarks are in line with the viewpoint that nonholonomic systems bear many
similarities with holonomic systems, as pointed out  by
Prof. \'Sniatycki in this meeting.}. 

It is common knowledge that 
  constraints   count
in double holonomic mechanics.   The Lagrangian vectorfield  is a spray: a
restriction on
$\dot{q}$ affects its ``twin brother'' in $T(TQ)$. This  suggests that
constraints  also count in double for nonholonomic systems. Using  the identification $\partial/\partial m_{\alpha} \equiv \epsilon_{\alpha}$ (a vertical vector),
it follows that
\begin{equation}
\partial H/\partial m_{\alpha} = dH_{|(q,m)}(\epsilon_{\alpha}) = \epsilon_{\alpha}(q)(\partial H/ \partial
p) = \epsilon_{\alpha}(q) (\dot{q}) \,\,.
\end{equation}
Here we consider  $\partial H/ \partial p \in T_qQ \equiv (T^*_qQ)^*
\equiv
(T_{p_q} T^*Q)^* $.
Therefore,   condition (i)   is a consequence of the
 constraint 
$\dot{q} \in E_q$. This condition ``does it
twice'', in the
construction of the reduced space $P$ and  in the projection to $Q$.
The vanishing middle rows and columns   in (\ref{prepoisson}) means the almost Poisson bracket of $\epsilon_{\alpha}$  with any differential $\xi  \in
T^*_{p_q} (T^*Q)$ is  zero. We call $\epsilon_{\alpha}$ a  
 {\it almost Casimir}.
 As for an ordinary Casimir in Poisson geometry, this implies that
$\epsilon_{\alpha} (X) = 0$ for any constrained
vectorfield $X$,  equivalent to the
statement that $\pi_*(X) \in \mathcal{E}$.
Any exact combination of the
$\epsilon_{\alpha}$'s will produce a bona fide  Casimir
function on $P$. Actually, these will be functions
on $Q$, because the $\epsilon_{\alpha}$ are {\it  basic}
differentials. Since we are interested in strictly
nonholonomic systems,  we may assume that
no exact combinations   exist \footnote{This does not rule out ``gauge conservation laws'', see .eg. \cite{BGM}, which appear whenever a combination of group action generators satisfies the constraints.}.  

We finish with a spiritual observation, which we hope proper, both in terms of mathematics and religion as well. 
Mathematicians use a universal handwaving gesture to represent a riemannian manifold, through a moving frame attached to it. A similar gesture to represent a symplectic manifold is in order. We believe that such gesture 
(``mudra'') may be found in Buddhism: Siddhartha's right hand explores the earth (a lagrangian submanifold), 
the left hand explores the spiritual fiber (another lagrangian submanifold). In so doing, the earthly hand acquires a spiritual component.



\newpage 

\begin{thebibliography}{99}

\bibitem{Arnold} Arnold, V.I. Arnold, V.V. Kozlov, and A.I. Neishtadt, {\em Dynamical Systems III\/},
Encyclopaedia of Mathematical Sciences, vol. ~3, Springer, New York 1988.

\bibitem{B}  L. Bates, Examples of singular nonholonomic reduction, {\it Rep. Math. Phys.}, {\bf 42}(1/2), 231-247 (1998).   

\bibitem{Bates} L. Bates and R. Cushman, What is a completely integrable nonholonomic dynamical system?,
{\em Rep.\ Math. \ Phys.\/} {\bf 44}(1/2), 29--35 (1999).  

\bibitem{BGM}  L. Bates, H. Graumann, C. MacDonnell, Examples of gauge conservation laws in nonholonomic systems, 
{\it Rep. Math. Phys.}, {\bf 37}(33), 295-308 (1996).  

\bibitem{Cartan}  E. Cartan,  Sur la repres\'entation
g\'eom\'etrique des syst\`emes mat\'eriels non holonomes, Proc. Int. Congr. Math., vol. {\bf 4}, Bologna,  253--261 (1928).

\bibitem{Cantrijn}  F.Cantrijn,  M. de Le\'on,  D.  Mart{\'\i}n de Diego, 
On almost-Poisson structures in nonholomic mechanics, {\em
Nonlinearity} {\bf 12 },   721--737 (1999). 

\bibitem{E} L. Euler, Du mouvement de rotation des corps solides autour d'un axe variable, 
{\it Mem. de l'acad. sci. Berlin}, {\bf 14}, 154-193 (1758). 

\bibitem{Koiller} J. Koiller, Reduction of some nonholonomic systems with symmetry,
{\em Arch.\ Rational Mech.\ Anal.\/}{\bf 118}, 113--148 (1992). 

\bibitem{JP} J. Koiller, P.R. Rodriguez, P. Pitanga, Nonholonomic connections following Elie Cartan, 
{\it An. Acad. Bras. Cienc.}, {\bf 73}(2), 165-190 (2001). 

\bibitem{RiosKoiller}  P. de M. Rios and J. Koiller, Non-holonomic systems with symmetry allowing a conformally symplectic reduction, 
in Proceedings HAMSYS 2001, to appear. 

\bibitem{SCS} D.J. Saunders, F. Cantrijn, W. Sarlet, Regularity aspects and hamiltonization of nonholonomic systems, {\it J. Phys. A} {\bf 32}, 6869-6890 (1999). 

\bibitem{SM} A.J. van der Schaft and B.M. Maschke,  On the hamiltonian
formulation of nonholonomic mechanical systems, {\em Rep.\ Math. \ Phys.\/} {\bf 34}(2), 225--233 (1994).  

\bibitem{Weinstein} A. Weinstein,   Sophus Lie and symplectic geometry, {\em Expositiones mathematicae }{\bf 1}, 95--96 (1983).

\end{thebibliography}
\end{document}